\documentclass[prl,aps,twocolumn,floats,showpacs]{revtex4}

\usepackage{graphicx}
\usepackage{psfrag}

\begin{document}

\title{Bell's theorem for general N-qubit states}
\author{Marek \.{Z}ukowski$^1$ and \v Caslav Brukner$^2$}
\affiliation{$^1$Instytut Fizyki Teoretycznej i Astrofizyki Uniwersytet Gda\'nski, PL-80-952 Gda\'nsk, Poland\\
$^2$Institut f\"ur Experimentalphysik, Universit\"at Wien, Boltzmanngasse 5, A--1090 Wien, Austria}


\begin{abstract}
We derive a single general Bell inequality which is a {\it necessary} and
{\it sufficient} condition for the correlation
function for N particles to be describable in a local and
realistic picture, for the case in which measurements on each
particle can be chosen between two arbitrary dichotomic
observables. We also derive a {\it necessary} and {\it sufficient}
condition for an arbitrary N-qubit mixed state to violate this
inequality. This condition is a generalization and reformulation
of the Horodeccy family condition for two qubits.
\end{abstract}


\maketitle

Local realism imposes constraints on statistical correlations of
measurements on multiparticle systems. They are in the form of
Bell-type inequalities
\cite{bell,chsh,ghz,mermin,ardehali,belinskii,PERESBELL,svozil}.
In a realistic theory the measurement results are determined by
"hidden" properties the particles carry prior to and independent
of observation. In a local realistic theory the results obtained
at one location are independent of any measurements, or actions,
performed at space-like separation. Quantum mechanics predicts
violation of these constraints. This is known as Bell's theorem
\cite{bell} .

However the problems a) what are the most general constraints on
correlations imposed by local realism, and b) which quantum states
violate these constraints, are still open. The latter has been
solved in general only in the case of two particles in pure states
\cite{gisin,gisinperes} and for two-qubit mixed states
\cite{3horodecki}. Only recently bounds for local realistic
description of a higher-dimensional system have been found in some
simple cases \cite{kaszlikowski,kaszlikowski1,collins}.

Here the answer to the two long-standing questions (a) and (b)
is presented  for the case of a standard Bell type experiment on N qubits. By a
standard Bell experiment we mean, one in which each local observer
is given a choice between two dichotomic observables. We first
derive a {\it single} general Bell inequality that summarizes all
possible local realistic constraints on the correlation functions
for a N-particle system.  From this inequality one obtains as
corollaries the Clauser-Horne-Shimony-Holt (CHSH) inequality
\cite{chsh} for two-particle systems and the
Mermin-Ardehali-Belinskii-Klyshko (MABK) inequalities for N
particles \cite{mermin,ardehali,belinskii}. We show that the
correlation functions in a standard Bell experiment can be
described by a local realistic model if and only if the general
Bell inequality is satisfied. Therefore the general Bell
inequality is a {\it sufficient} and {\it necessary} condition for
correlation functions, in such an experiment, to be describable
within a local realistic model. We also find a {\it necessary} and
{\it sufficient} condition for correlation functions for N qubits
in an arbitrary (mixed) quantum state to violate the general Bell
inequality in direct measurements. This condition is
generalization and reformulation of the one given by the Horodeccy
family \cite{3horodecki} for two qubits.

These results are not only of importance from the fundamental
point of view, but also as a research towards identifying ultimate
resources for quantum information processing. Recently it was
shown \cite{ScaraniPRL}, that there is a direct link between the
security of the quantum communication protocols, and the violation
of Bell inequalities.

We shall now derive the general Bell
inequality. Consider N observers and allow each of them to choose
between two dichotomic observables, determined by some local
parameters denoted here $\vec{n}_1$ and $\vec{n}_2$. We choose
such a notation of the parameters for brevity; of course each
observer can choose independently two arbitrary directions. The
assumption of local realism implies existence of two numbers
$A_j(\vec{n}_1)$ and $A_j(\vec{n}_2)$ each taking values +1 or -1,
which describe the predetermined result of a measurement by the
$j$-th observer of the observable defined by $\vec{n}_1$  and
$\vec{n}_2$, respectively (we do not discuss stochastic hidden
variable models, as they always can be constructed from underlying
deterministic ones). In a specific run of the experiment the
correlations between all $N$ observations can be represented by the
product $\prod_{j=1}^N A_j(\vec{n}_{k_j})$, with $k_j=1,2$. The
correlation function, in the case of a local realistic theory, is
then the average over many runs of the experiment
\begin{equation}
E({k_1},...,{k_N})=\big\langle  \prod_{j=1}^N
A_j(\vec{n}_{k_j})\big\rangle_{avg}. \label{corrfunction}
\end{equation}

The following algebraic identity holds for the predetermined results:
\begin{equation}
\hspace{-0.15cm}  \sum_{s_1,..,s_N =\pm1} \hspace{-0.2cm}  \!
S(s_1,...,s_N)\prod_{j=1}^N [ A_j(\vec{n}_{1}) + s_j
A_j(\vec{n}_{2})]=\pm2^N, \label{INEQ2}
\end{equation}
where $S(s_1,...,s_N)$ stands for an arbitrary function of the
summation indices $s_1,...,s_N\in \{-1,1\}$, such that its values
are only $\pm1$, i.e. $S(s_1,...,s_N)=\pm1$. To prove this
identity, note, that, since $A_j(\vec{n})=\pm1$, for each observer
$j$ one has either $|A_j(\vec{n}_1)+A_j(\vec{n}_2)|=0$ and
$|A_j(\vec{n}_1)-A_j(\vec{n}_2)|=2$, or the other way around.
Therefore, for all sign sequences of $s_1,...,s_N$ the product
$\prod_{j=1}^N [A_j(\vec{n}_{1}) + s_j A_j(\vec{n}_{2})]$ vanishes
except for just {\it one} sign sequence, for which it is $\pm2^N$.
If one adds up all such $2^N$ products, with an arbitrary sign in
front of each of them, the sum is always  equal to the value of
the only non-vanishing term, i.e., it is $\pm 2^N$.

After averaging the expression (\ref{INEQ2}) over the ensemble of
the runs of the experiment (compare Eq. (\ref{corrfunction})) one
obtains the following set of Bell inequalities
\begin{equation}
|\hspace{-2mm} \sum_{s_1,...,s_N \atop = -1,1} \hspace{-2mm}
S(s_1,...,s_N) \hspace{-3mm} \sum_{k_1,...,k_N \atop = 1,2}
\hspace{-2mm} s^{k_1-1}_1.. s^{k_N-1}_N E({k_1},...,{k_N})| \leq
2^N. \label{allbellineq}
\end{equation}

Since there are $2^{2^N}$ different functions $S(s_1,...,s_N)$,
the inequalities (\ref{allbellineq}) represent a set of $2^{2^N}$
Bell inequalities for the correlation functions. Many of these
inequalities are trivial. E.g., when the choice for the function
is $S(s_1,...,s_N)=1$ for all arguments, we get the condition
$E(1,1,...1)\leq1$. Specific other choices give non-trivial
inequalities. For example, for $S(s_1,...,s_N) =\sqrt{2}
\cos(-\frac{\pi}{4}+(s_1+...+s_N-N)\frac{\pi}{4})$ one recovers
the MABK inequalities, in the form derived by Belinskii and
Klyshko \cite{belinskii}. Specifically, for $N=2$, the well known
CHSH inequality \cite{chsh} follows. For $N=3$, one obtains the
inequality
\begin{equation}
|E(1,2,2)+E(2,1,2)+E(2,2,1)-E(1,1,1)| \leq 2. \label{mermin}
\end{equation}
Inequalities, like the one above, with the minus sign at a
different location, and/or measurements $1$ and $2$ permuted, form
together an equivalence class.

The full set of all $2^{2^N}$ inequalities (\ref{allbellineq}) is
equivalent to the {\it single} general Bell inequality
\cite{weinfurter,zukowskibrukner,werner}
\begin{equation}
\hspace{-2mm} \sum_{s_1,...,s_N \atop =-1,1} \hspace{-1mm} |
\hspace{-1mm} \sum_{k_1,...,k_N \atop = 1,2} s^{k_1-1}_1 ...
s^{k_N-1}_N \mbox{ } E({k_1},...,{k_N})|\leq 2^N.
\label{thebellineq}
\end{equation}
The equivalence of (\ref{thebellineq}) and (\ref{allbellineq}) is
evident, once one  recalls that, for real numbers one has $|a+
b|\leq c$ and $|a-b|\leq c$  if and only if $|a|+|b|\leq c$, and
writes down a generalization of this property to sequences of an
arbitrary length.

Thus far we have shown that when a local realistic model exists,
the general Bell inequality (\ref{thebellineq}) follows. The
converse is also true: whenever inequality (\ref{thebellineq})
holds one can construct a local realistic model for the
correlation function, in the case of a standard Bell experiment.
This establishes the general Bell inequality (\ref{thebellineq})
presented above as a {\it necessary} and {\it sufficient}
condition for local realistic description of $N$ particle
correlation functions in standard Bell-type experiments.
 This is why one can claim that the set of Bell
inequalities (\ref{allbellineq}) is {\it complete}.

The proof of the sufficiency of condition (\ref{thebellineq}) will
be done in a constructive way. A local realistic theory must
ascribe certain probabilities to every possible set of
predetermined local results. Just like if the local measuring
stations were receiving instructions, what should be the
measurement results for (here) two possible settings of the local
apparatus.

One can ascribe to the set of predetermined local results, which
satisfy the following conditions
$A_j(\vec{n}_1)=s_jA_j(\vec{n}_2)$, the hidden probability
\begin{equation}
p(s_1,..,s_N) \!= \!\frac{1}{2^N}| \hspace{-0.2cm}
\sum_{k_1,..,k_N} \hspace{-0.2cm} s^{k_1-1}_1 .. s^{k_N-1}_N
E({k_1},...,{k_N})|,
\end{equation}
and one can demand that the product $\prod_{j=1}^N
A_j(\vec{n}_{1})$ has the same sign as that of the expression
inside of the modulus defining the $p(s_1,...,s_N)$. In this way every
{\it definite} set of local realistic values is ascribed a {\it
unique} global hidden probability. However, if the inequality
(\ref{thebellineq}) is not saturated the probabilities add up to
less than 1. In such a case, the "missing" probability is ascribed
to an arbitrary model of local realistic noise (e.g., for which
all possible products of local results enter with equal weights).
The overall contribution of such a noise term to the correlation
function is nil. In this way we obtain a local realistic model of
a certain correlation function.

However, one should check whether this construction indeed
produces the model for the correlation function for the set of
settings that enter inequality (\ref{thebellineq}), that is for
$E(k_1,...,k_N)$. For simplicity take $N=2$. One can build a
``vector" $(E(1,1),E(1,2), E(2,1),E(2,2))$ out of the values of
the correlation function. The expansion coefficients of this
``vector" in terms of the four orthogonal basis vectors
$(1,s_1,s_2, s_1s_2)$ (recall, that $s_1,s_2 \in \{-1,1\}$) are
equal to the expressions within the moduli entering inequality
(\ref{thebellineq}). By the construction shown above the local
realistic model for $N=2$ gives the following ``vector"
\begin{eqnarray}
\lefteqn{(E_{LR}(1,1),E_{LR}(1,2), E_{LR}(2,1),E_{LR}(2,2)) }
\\ & \hspace{0.5cm} &= \frac{1}{4} \sum_{s_1,s_2} \big[\sum_{k_1,k_2}
s^{k_1-1}_1 s^{k_2-1}_2 E({k_1},{k_2})\big] (1,s_1,s_2, s_1s_2).
\nonumber
\end{eqnarray}
Since the vector built out of the correlation function values and
its local realistic counterpart have the {\it same expansion
coefficients} in the basis, they are equal. Thus, the sufficiency of
(\ref{thebellineq}) as a condition
for the existence of a local realistic model is proven.
The generalization to an arbitrary $N$ is obvious.

Quantum mechanical predictions can violate the inequality
(\ref{thebellineq}). Simply, if a MABK inequality is violated,
then the general inequality, which also includes the MABK
inequalities, is violated too. However, the converse statement is
not always true. The new inequality is more restrictive. In the
problem of identifying quantum states of highly nonclassical
traits, it is important to find the class of quantum states, which
are not describable by local realistic models. We will now derive
the necessary and sufficient condition for an arbitrary (pure or
mixed) quantum state to violate the general Bell inequality
(\ref{thebellineq}).

An arbitrary mixed
state of $N$ qubits can be written down as
\begin{equation}
\rho=\frac{1}{2^N} \sum_{x_1,...,x_N=0}^{3} T_{x_1...x_N} \mbox{ }
\sigma^1_{x_1} \otimes ... \otimes \sigma^N_{x_N}, \label{state}
\end{equation}
where $\sigma^{j}_0$ is the identity operator in the Hilbert space
of qubit $j$, and $\sigma^{j}_{x_j}$ are the Pauli operators for
three orthogonal directions $x_j=1,2,3$. The set of real
coefficients $T_{x_1...x_N}$, with $x_j=1,2,3$ forms  the
so-called  correlation tensor $\hat{T}$. The correlation tensor
fully defines the N-qubit correlation function:
\begin{eqnarray} \hspace{-0.6cm} 
E_{QM}(k_1,...,k_N)&=& \mbox{Tr}[\rho \mbox{ } (
\vec{n}_{k_1}\cdot \vec{\sigma}\otimes ... \otimes \vec{n}_{k_N}\cdot\vec{\sigma}
)] \label{corrquantum} \\ &=&
\hspace{-0.5cm}\sum_{x_1,...,x_n=1}^3 \hspace{-0.3cm}
T_{x_1...x_N} (\vec{n}_{k_1})_{x_1} ... (\vec{n}_{k_N})_{x_N},
\label{EXPR}
\end{eqnarray}
where $(\vec{n}_{k_j})_{x_j}$  are the three Cartesian components
of the vector $\vec{n}_{k_j}$. For convenience we shall write down
the last expression (\ref{EXPR}) in a more compact way as
$\langle\hat{T},\vec{n}_{k_1} \otimes ... \otimes
\vec{n}_{k_N}\rangle$, where $\langle ...,...\rangle$ denotes the
scalar product in $R^{3N}$.

We now insert the quantum correlation function
$E_{QM}(k_1,...,k_N)$ into the Bell inequality
(\ref{thebellineq}), and obtain
\begin{equation}
\sum_{s_1,...,s_N \atop = -1,1} \hspace{-1mm}
|\langle\hat{T},\sum_{k_1=1}^2 s^{k_1-1}_{1} \vec{n}_{k_1} \otimes
... \otimes \sum_{k_N=1}^2 s^{k_N-1}_N \vec{n}_{k_N}\rangle| \leq
2^N. \label{statojaradim}
\end{equation}
This inequality can be simplified. For each observer there always
exist two mutually orthogonal unit vectors $\vec{a}^j_1$ and
$\vec{a}^j_2$ and the angle $\alpha_j$ such that $\sum_{k_j=1}^2
\vec{n}_{k_j}\! =\! 2 \vec{a}^j_1 \cos(\alpha_j + \frac{\pi}{2})$
and $\sum_{k_j=1}^2 (-1)^{k_j} \vec{n}_{k_j} \!=\! 2 \vec{a}^j_2
\cos(\alpha_j + \pi)$. Using the notation $c^j_{x_j}\! =
\!\cos(\alpha_j + x_j \frac{\pi}{2})$, one can write the
inequality (\ref{statojaradim}) as
\begin{equation}
\sum_{x_1,...,x_N=1}^{2} |c^1_{x_1} ... c^N_{x_N} \langle \hat{T},
\vec{a}^1_{x_1} \otimes ... \otimes \vec{a}^N_{x_N}\rangle| \leq
1.
\end{equation}
One can transform this inequality into
\begin{equation}
\sum_{x_1,...,x_N= 1}^{2} c^1_{x_1} ... c^N_{x_N} |T_{x_1...x_N}|
\leq 1 \label{necsufcondition}
\end{equation}
where $T_{x_1...x_N}$ is now a component of the tensor $\hat{T}$
in a {\it new} set of local coordinate systems, which among their
basis vectors have $\vec{a}^j_1$ and $\vec{a}^j_2$. The two
vectors serve as the unit vectors which define, say, the local
directions $x$ and $y$. The values of $c^j_{x_j}$ enter
(\ref{necsufcondition}) directly, not as moduli, because without
this constraint the maximal  value of the left hand side does not
change.

We conclude from the above reasoning that the {\it necessary} and
{\it sufficient} condition for an arbitrary N-qubit state to
satisfy the general Bell inequality (\ref{thebellineq}) can be put
in the following way. {\it The correlations between the
measurements on $N$ qubits satisfy inequality (\ref{thebellineq}) if
and only if in {\it any} set of local coordinate system of N
observers, and for any set of unit vectors
$\vec{c}^j=(c^j_1,c^j_2)$ one has}
\begin{equation}
T^{mod}_{c_1...c_N} \equiv \sum_{x_1,...,x_N= 1}^{2} c^1_{x_1} ...
c^N_{x_N} \mbox{ } |T_{x_1...x_N}| \leq 1. \label{mod}
\end{equation}

Let us give a geometric interpretation of (\ref{mod}). Suppose one
replaces the components of the correlation tensor $T_{x_1...x_N}$
by their moduli $|T_{x_1...x_N}|$, and builds of such moduli a new
tensor $\hat{T}^{mod}$. Suppose moreover one transforms this
modified tensor into a new set of local coordinate systems, each
of which is obtained from the old one by a rotation within the
plane spanned by axes 1 and 2 of the initial coordinates. If this
new tensor satisfies constraint (\ref{mod}) for an arbitrary
choice of the initial set of local coordinate systems, then, and
only then, a local realistic description of correlation function
is possible, in the case of {\it any} standard Bell experiment.

In other words, $T^{mod}_{c_1...c_N}$ is a  component of $T^{mod}$
along directions defined by the unit vectors $\vec{c}^j$,
$j=1,...,N$. If the condition (\ref{mod}) holds, then the
transformed components $\hat{T}^{mod}_{c_1,...c_N}$ do not have
values larger than 1. Only then, they can describe products of
local results, which are only of the values $\pm 1$, like is for
any correlation tensor. One therefore can express the condition
(\ref{mod}) as follows: within local realistic description
$\hat{T}^{mod}$ is also a possible correlation tensor. This bears
a similarity with the Peres \cite{peressep} necessary condition
for separability (a partially transposed density matrix is a
possible density matrix). Note that (\ref{mod}) could also be put
in yet another way: arbitrary changes of the signs of some of the
coordinates of $\hat{T}$ still leave it as a possible correlation
tensor.

By applying the Cauchy inequality to the middle term of expression
(\ref{mod}) one obtains directly the following useful and simple
{\it sufficient} condition for local realistic description of the
correlation functions for $N$ qubits. If in {\it any} set of local
coordinate systems of $N$ observers
\begin{equation}
\sum_{x_1,...,x_N = 1}^{2} T^2_{x_1...x_N} \leq 1, \label{yeah}
\end{equation}
then the correlations between the measurements on $N$ qubits satisfy the general inequality
(\ref{thebellineq}).

By performing rotations in the planes defined by directions $1$
and $2$ of each of  the $N$ observers one can vary the values of the
elements of the correlation tensor, but these variations do not
change the left-hand side of inequality (\ref{yeah}). In this way,
one can find local coordinate systems for which some of the
correlation tensor elements vanish. Thus the criterion
(\ref{yeah}) can involve a smaller number of them (compare the
three qubit case in Ref. \cite{scarani}).

There are special situations for which (\ref{yeah}) turns out to be both
the necessary and sufficient condition for correlation functions
to satisfy the general Bell inequality (\ref{thebellineq}).
Formally this arises whenever the two ''vectors"
$(c^1_1...c^N_1,...,c^1_2...c^N_2)$ and $(|T_{11...1}|,...,
|T_{22...2}|)$ in (\ref{necsufcondition}) are parallel. Only then,
since the first vector has a unit norm, the expression on the left
hand side of (\ref{necsufcondition}) reaches the one on the left
hand side of (\ref{yeah}) and thus the conditions
(\ref{necsufcondition}) and (\ref{yeah}) are equivalent ones.

Let us consider two examples of application of our results. We
first study an arbitrary two-qubit state to recover the Horodeccy
condition \cite{3horodecki}. In this case, since the two
''vectors" $(|T_{11}|, |T_{12}|,|T_{21}|,|T_{22}|)$ and
$(c^1_1c^2_1,c^1_1c^2_2,c^1_2c^2_1,c^1_2c^2_2)$ in
(\ref{necsufcondition}) can be made parallel by a suitable choice
of free parameters, (\ref{yeah}) is the necessary and sufficient
condition for the violation of local realism. Two of the
parameters come from the arbitrariness in selecting the two local
coordinate systems (i.e. they come from arbitrary transformation
of the correlation tensor by rotations within $1$-$2$ planes of
each of the two local coordinate systems). Two more parameters are
the two angles $\alpha_j$ which define the second ``vector".
Therefore the condition reads $T_{11}^2+T_{12}^2+T_{21}^2 +
T_{22}^2 \leq 1$. In addition one can always find local coordinate
systems such that $T_{12}=T_{21}=0$ and our condition transforms
into $T^2_{11}+T^2_{22} \leq 1$. This is equivalent to the
Horodeccy condition  \cite{3horodecki}, as $T^2_{11}$ and
$T^2_{22}$ for the diagonalized correlation tensor are equal to
two eigenvalues of the matrix $\hat{T}^T\hat{T}$, where
$\hat{T}^T$ is the transposed $\hat{T}$.

As another example, we consider the Werner states. Such states
have the form $ \rho_W \!=\! V |\psi_{GHZ} \rangle \langle
\psi_{GHZ}|\! +\! (1\!-\!V) \rho_{noise},$ where $ |\psi_{GHZ}
\rangle \!=\! \frac{1}{\sqrt{2}} (|0\rangle_1...|0\rangle_N \!+\!
|1\rangle_1...|1\rangle_N)$ is the maximally entangled (GHZ) state \cite{ghz}
and $\rho_{noise} = I/2^N$ is the completely mixed state. Here the
weight $V$ of the GHZ-state can operationally be interpreted as
the interferometric contrast observed in a multi-particle
correlation experiment. The nonvanishing components of the
correlation tensor in the $xy$ planes for the Werner state are:
the components which contain an even number of $y$'s and
$T_{xx...x}$. There are alltogether $2^{N-1}$ of them. Their
values are either $+V$ or $-V$. Since again the two ``vectors" in
(\ref{necsufcondition}) can be made parallel, (\ref{yeah}) is the
necessary and sufficient condition for the violation of local
realism. Indeed, if one rotates all but one local coordinate
systems by $45^o$, then all $2^N$ components of the ``vector"
$(|T_{11...1}|,..., |T_{22...2}|)$ become equal to $V/\sqrt{2}$.
Furthermore, if one chooses all $\alpha_j$ equal to $-\pi/4$, the
unit ``vector" $(c^1_1...c^N_1,...,c^1_2...c^N_2)$ has all its
components equal to $1/\sqrt{2^{N}}$. Therefore, the two
``vectors" are parallel. Thus, using criterion (\ref{yeah}) we
conclude that the correlation functions for the Werner state
definitely cannot be described by local realism if and only if $V
> 1/\sqrt{2^{N-1}}$.

More applications of the formalism, leading to some unexpected
results, are given in \cite{marek+}. There a family of pure
entangled states is found, which {\it do not violate any} Bell
inequality for correlation functions, for the standard Bell
experiment.

It will be interesting to see
generalizations of the criteria for violation of local realism to
the cases of higher-dimensional systems than qubits and to more
measurement choices for each observer than two. One can expect in
such cases even stronger restrictions for the local realistic
description \cite{zukowski}.

M.\.{Z}. acknowledges KBN grant No. 5 P03B 088 20. {\v C}.B. is
supported by the Austrian  FWF project F1506, and by the QIPC
program of the EU. The work is a part of the Austrian-Polish
program {\em Quantum Communication and Quantum Information}. Both
authors acknowledge discussions with Anton Zeilinger.

\end{document}